\documentclass[preprint]{revtex4-2}
\usepackage{mathtools}
\usepackage{amsmath,bm}
\usepackage{amssymb}
\usepackage{subcaption}
\usepackage{setspace}
\usepackage{dcolumn}
\usepackage{color}
\usepackage{physics}
\usepackage[hidelinks]{hyperref}
\usepackage{bm}
\usepackage{multirow}
\usepackage{booktabs} 

\begin{document}

\title{Hyperbolic Space Spectral Characteristics in a Network of Mechanical Linkages}

\author{Nicholas H. Patino, Curtis Rasmussen, Massimo Ruzzene}
\affiliation{Department of Mechanical Engineering, University of Colorado Boulder, Boulder, Colorado 80309, USA}
\date{\today}

\begin{abstract}
We investigate the dynamic properties of elastic lattices defined by tessellations of a curved hyperbolic space. The lattices are obtained by projecting nodes of a regular hyperbolic tessellation onto a flat disk and then connecting those sites with simple linkages. Numerical and experimental investigations illustrate how their vibrational spectral properties are characterized by a high density of modes that are localized at the domain boundaries. Such properties govern the propagation of waves induced by broadband inputs. This suggests the potential for applications seeking the protection of bulk media from boundary-incident perturbations. We uncover the boundary-dominated nature of an exemplary hyperbolic lattice through the evaluation and analysis of its integrated density of states and vibrational spectrum. The dynamics of the lattice are also contextualized by comparing them with those of continuous disks characterized by Euclidean and hyperbolic property distributions, which confirms that the lattice retains the boundary-dominated spectrum observed in the hyperbolic plane. We then numerically investigate the response of the lattice to transient pulses incident on the boundary and find that edge-confined wave propagation occurs. The modal and transient pulse propagation behavior of the lattice is experimentally validated in a milled aluminum sample. By leveraging hyperbolic geometry, our mechanical lattice ushers in a novel class of mechanical metamaterials with boundary-dominated wave phenomena reminiscent of topologically protected systems suitable for applications in advanced wave control.

\end{abstract} 

\maketitle

\section{Introduction}

It is known that waves can be controlled and steered by modulating the material properties of a wave-bearing medium in space or time. However, it is often difficult in practice to exert the amount of control necessary on the material properties to achieve the desired wave properties. A passive approach to wave control is to curve the underlying geometric space hosting the waves~\cite{zhu2018elastic,lee2021singular,mazzotti2022elastic,schultheiss2010optics,bekenstein2017control,ratcliffe1994foundations,helgason1994geometric}. In this case, the material can remain a simple homogeneous material while the wave propagation is controlled by the curvature of the medium. Curved materials are ubiquitous in the natural world where they are known to induce collective cellular wave phenomena in biological tissues~\cite{brandstatter2023curvature,schamberger2023curvature}, scatter or focus waves on surfaces~\cite{mitchell2014lenses,rulf1969rayleigh,la2019curvilinear}, and allow for topologically protected waves in acoustic systems as well as Earth's atmosphere and oceans~\cite{shankar2017topological,delplace2017topological}.

A curved space of note is the hyperbolic plane which is characterized by a constant negative curvature over its entire domain. This curious space challenges our geometric intuition by rejecting Euclid's parallel postulate allowing for peculiar mathematical implications reported as early as the 18th century \cite{saccheri1986euclides}. Studies in the 19th century then developed the canonical differential geometry of hyperbolic space~\cite{beltrami1868saggio,killing1879ueber,bolyai1832appendix,klein1872ueber}, while more recent investigations have developed its geometric group theory~\cite{series1981infinite,beardon1983geometry,woess1987context,plotnick1987growth,corlette1992archimedean,lubotzky1996free,katok1992fuchsian} and classical lattice dynamics~\cite{comtet1987landau,balazs1986chaos,rietman1992ising,monthus1996random,aurich1996trace,bunimovich1991statistical,ueda2007corner,baek2009percolation,baek2008diffusion}. In the last few years, hyperbolic lattices have been studied as platforms for informing the connection between quantum mechanics and general relativity through simulations of quantum lattices embedded in hyperbolic space~\cite{bienias2022circuit,boettcher2020quantum,lenggenhager2022simulating,kollar2019hyperbolic}. In another vein of recent research, the theory of topological insulators, previously only considered in flat, Euclidean space, has been investigated in hyperbolic space, introducing new platforms for topologically protected systems~\cite{cheng2022band,liu2022chern,urwyler2022hyperbolic,chen2023hyperbolic,lux2023converging,yu2020topological,zhang2022observation,zhang2023hyperbolic,fleury2023anomalous}. These active areas of research have been supported by advances in the mathematical descriptions of waves in hyperbolic space, including hyperbolic crystallography and band structure~\cite{crystallographyhyp,lux2023converging,maciejko2022automorphic,maciejko2021hyperbolic,attar2022selberg,stegmaier2022universality}. Due to the inherent difficulties in creating practical structures with constant negative curvature, work has been done to mimic hyperbolic space in flat domains. Of note is recent work with circuit networks which has shown how dynamics in a hyperbolic space can be realized in tabletop experiments~\cite{lenggenhager2022simulating,boettcher2020quantum}. For electrical circuits, where the lengths of coupling wires are orders smaller than a wavelength, the resonators, placed at lattice sites, are effectively connected with equivalent couplings regardless individual wire lengths. Circuit networks have since been leveraged to experimentally explore topological phases in hyperbolic space~\cite{zhang2022observation,zhang2023hyperbolic}. Recently, the exploration of topological wave phenomena in hyperbolic systems has been extended to electromagnetic waves with a notable study in a non-reciprocal scattering network hosting microwave frequencies~\cite{fleury2023anomalous} reporting robust chiral topological edge modes of Chern and anomalous types.

Previous work in mechanics include the theoretical study conducted in~\cite{ruzzene2021dynamics}, which highlights properties of hyperbolic tessellations with length-dependent interactions. In the present work, we expand these investigations to specifically focus on the analysis of spectral properties and the existence of high densities of states that are confined to the boundaries. This is a property that appears tightly connected to the generation of the tessellation, which produces approximants of hyperbolic distributions of material properties. Our experimentally-validated numerical simulations confirm the existence of a high density of boundary-dominated modes, and suggest that the considered lattice is a first illustration of hyperbolic space dynamics in a mechanical system.

Following this introduction, we begin by reviewing hyperbolic geometry and the Poincaré disk model of hyperbolic space used to design the proposed mechanical lattice. To illustrate our lattice's ability to simulate hyperbolic space, we also review the dynamics governing harmonic vibrations in continuous Euclidean and hyperbolic spaces. With these models established we show, by examining the integrated density of states of all three systems, that the lattice exhibits a boundary-dominated spectra similar to that of the hyperbolic plane. Next, the time-dependent response to boundary-incident pulses is numerically explored in connection to the previously investigated spectral properties. We find that edge-confined wave propagation occurs in the lattice, which we experimentally corroborate through laser Doppler vibrometry measurements of a milled aluminum sample. These findings demonstrate the promise of elastic hyperbolic lattices as a feasible class of metamaterials for advanced wave control with applications including vibration mitigation, vibration-based sensing, and waveguiding. The proposed design presents a simple platform with potential for advanced studies in hyperbolic wave phenomena, such as novel topological states and the realization of curved spaces with novel physical properties and wave guiding capabilities.

\section{Hyperbolic geometry on a flat surface}

\begin{figure}[!ht]
	\centering
	\includegraphics{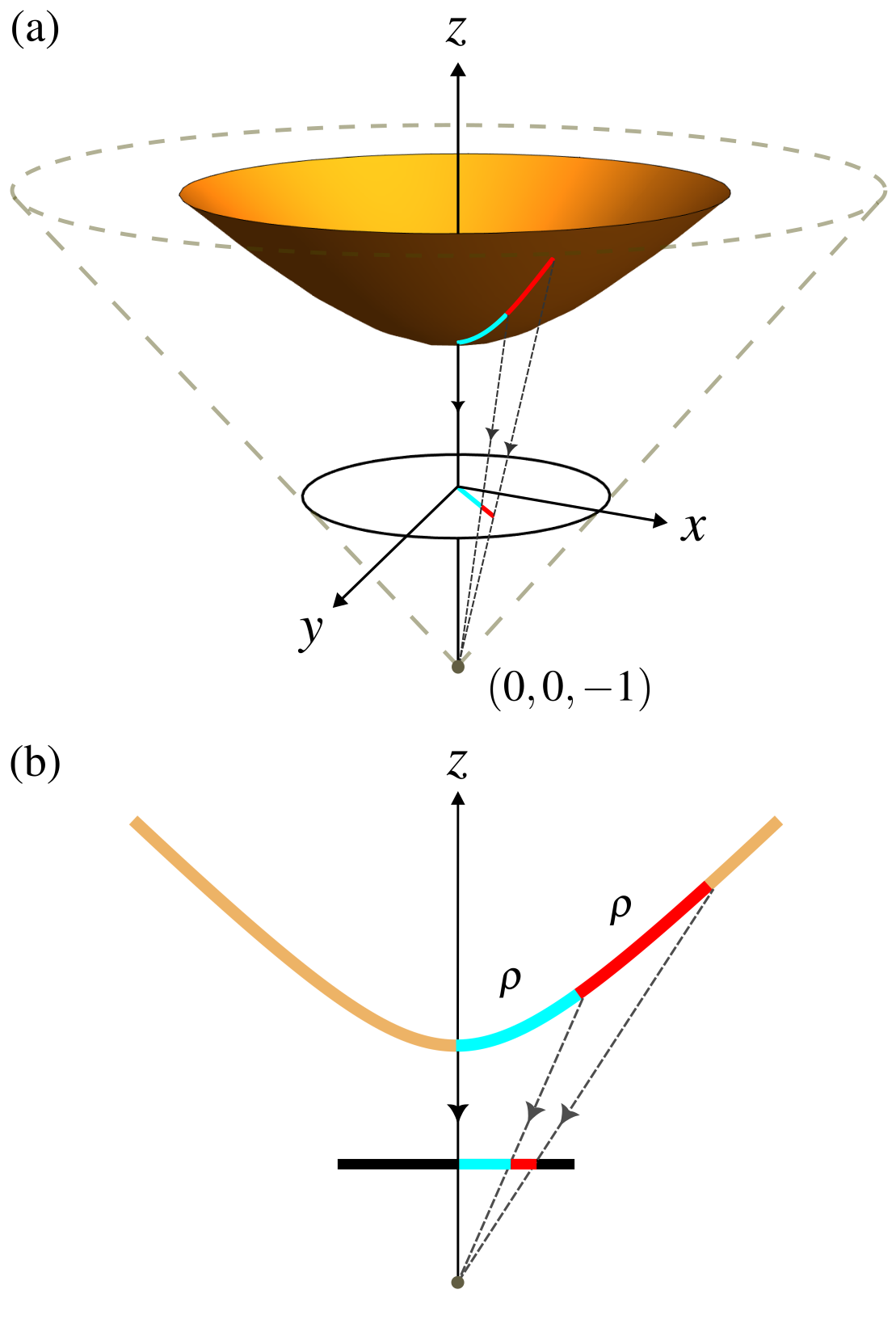}
	\caption{The Poincar\'e disk model of hyperbolic space as a projection of the hyperboloid model of hyperbolic space. (a) The Poincar\'e disk is obtained by projecting the hyperboloid (in orange) onto a disk of radius one through the point $(0,0,-1)$. Cyan and red line segments show two geodesics in both models, which are of equal length $\rho$ in hyperbolic space. (b) Cross-section of diagram in panel (a) illustrating the projection of hyperbolic geodesics from the hyperboloid to the disk, where compression of distances along the radius is seen.}
	\label{Fig1}
\end{figure}

The hyperbolic plane is a two-dimensional surface characterized by constant negative Gaussian curvature, in contrast to the surface of a sphere which has everywhere a constant positive Gaussian curvature. Hilbert's theorem states that the hyperbolic plane cannot be isometrically mapped to the 3D Euclidean space \cite{hilbert1970flachen,do2016differential}, meaning that there are no complete surfaces with constant negative curvature in Euclidean space which faithfully represent hyperbolic space. Thus, all models of hyperbolic space in subspaces of $\mathbb{R}^3$ necessarily distort lengths. One common representation of the hyperbolic plane is the positive sheet of a two-sheet hyperboloid which extends infinitely in a three-dimensional Euclidean space. This is the hyperboloid model of the hyperbolic plane, a portion of which is shown in orange in Fig. \ref{Fig1}. This model provides a visually intuitive curvature but exists in three spatial dimensions which limits manufacturing techniques from a homogeneous material and complicates experimental measurements. A more practical alternative is to project the hyperboloid to a compact, lower-dimensional domain, where it is easier to manufacture and study. Much like how cartographers project Earth’s positively curved surface to a flat map, the curved surface of the hyperboloid can be projected onto a flat unit disk with all projection lines meeting at point (0,0,-1) as shown in Fig. \ref{Fig1}a. This flat model of the hyperbolic plane, known as the Poincar\'e disk model~\cite{poincare1882memoire}, is a conformal mapping that locally preserves hyperbolic angles and thus the shape of hyperbolic polygons in a tessellation.

Each model of the hyperbolic plane is paired with a metric that defines distances in the hyperbolic plane as a function of distances in the model. We illustrate these metrics by drawing two arcs, in cyan and red, in Fig. \ref{Fig1}. These arcs, which are of equal length $\rho$ in hyperbolic space, are seen as arcs of unequal length on the hyperboloid and on the disk. In the former they appear to stretch away from the origin, whereas in the latter they appear to compress away from the origin. For the Poincar\'e disk, this compressive effect is mathematically reconciled by equipping the disk with the non-Euclidean Poincar\'e metric: ~\cite{beltrami1868saggio,poincare1882memoire2}

\begin{equation}
    d\rho^2 = -\frac{4}{K} \frac{dx^2+dy^2}{(1-r^2)^2},
    \label{metric}
\end{equation}

\noindent where the hyperbolic line element $d\rho$ is related to Euclidean distances through the Gaussian curvature $K$ and the radial coordinate $r=\sqrt{x^2+y^2}$. $K$ quantifies the curvature of the hyperbolic space and is commonly assigned the value $-4$ or $-1$~\cite{ attar2022selberg,lenggenhager2022simulating}.

\section{Wave dynamics on the Poincar\'e disk}

A direct result of the Poincar\'e metric in \ Eq.~\eqref{metric} is not only the distortion of distances on the disk but also the distortion of wave speeds. In this section, we illustrate how wave dynamics are distorted in the Poincar\'e disk by first considering the classical wave equation for an isotropic Euclidean space, given by

\begin{equation}
    \Delta_E u = \frac{1}{c_0^2} \frac{\partial^2 u}{\partial{t}^2},
    \label{Euclidean wave eq}
\end{equation}

\noindent where $\Delta_E$ denotes the Laplacian in Euclidean space, $u$ is the field variable, and $c_0$ is the characteristic wave speed.

In the Poincar\'e disk, we see that the spatial derivatives of the wave equation must be modified to account for Eq.~\eqref{metric}. The Laplacian generalizes to the Laplace-Beltrami operator~\cite{jost2008riemannian,balazs1986chaos,helgason1979differential}. Acting on a smooth function $\phi$, this operator is given by

\begin{equation}
       \Delta \phi = \frac{1}{\sqrt{|g|}} \sum_{i,j} \frac{\partial}{\partial{x}_i}(\sqrt{|g|}g^{ij}\frac{\partial \phi}{\partial{x}_j}),
    \label{laplace-beltrami}
\end{equation}

\noindent where $|g|$ is the determinant of the metric tensor $g_{ij}$ relative to coordinates $x_1$ and $x_2$ of a two-dimensional space and $g^{ij}$ is the inverse of $g_{ij}$, following Ricci calculus tensor notation. In Euclidean space, the metric tensor is given by the identity matrix, $g_{ij}=\delta_{ij}$, and Eq.~\eqref{laplace-beltrami} gives the Laplacian $\Delta_E=\frac{\partial^2}{\partial{x}^2}+\frac{\partial^2}{\partial{y}^2}$, where $(x,y)$ defines an ordered pair of Cartesian coordinates on the reference plane. The metric tensor for the Poincar\'e disk with curvature $K=-1$ is $g_{ij}=\frac{4}{(1-x^2-y^2)^2}\delta_{ij}$, leading to a Laplace-Beltrami operator, which upon comparison with Eq.~\eqref{Euclidean wave eq}, can be expressed as

\begin{equation}
    \Delta_H = \frac{(1-x^2-y^2)^2}{4} (\frac{\partial^2}{\partial{x}^2}+\frac{\partial^2}{\partial{y}^2}) = \frac{(1-r^2)^2}{4} \Delta_E,
    \label{hyperbolic laplacian}
\end{equation} 

\noindent where the subscript $H$ signifies action on functions defined on the hyperbolic plane. At the origin of the Poincar\'e disk, $r=0$, $\Delta_H$ reduces to $\frac{1}{4} \Delta_E$, which suggests that near this point the dynamics of waves on the Poincar\'e disk are governed by an equation $\frac{1}{4}\Delta_E u \approx \frac{1}{c_0^2} \frac{\partial^2 u}{\partial{t}^2}$. This corresponds to a Euclidean surface whose wave speed is reduced by a factor of 2. Moving away from the origin in a straight path, and with reference to Fig. \ref{Fig1}b, equidistant steps in the hyperbolic plane are increasingly compressed on the Poincar\'e disk, due to its radially-dependent metric. As we approach the outer rim at $r=1$, we approach an infinite compression of hyperbolic space and the step size becomes vanishingly small. This geometric feature appears in the wave equation through the radially-dependent spatial operator, which becomes the zero operator in the limit $r\rightarrow1$. Equation~ \eqref{hyperbolic laplacian} suggests that Poincar\'e disk wave dynamics could be achieved in a flat circular system where the wave speed is a function of $r$. This can be seen by considering the wave equation on the Poincaré disk,

\begin{equation}
    \Delta_H u = \frac{(1-r^2)^2}{4} \Delta_E u = \frac{1}{c_0^2} \frac{\partial^2 u}{\partial{t}^2},
    \label{hyperbolic wave eq}
\end{equation}
which is formally equivalent to

\begin{equation}
    \Delta_E u = \frac{1}{c(r)^2} \frac{\partial^2 u}{\partial{t}^2},
    \label{wave eq radial wave speed}
\end{equation}
and corresponds to a system whose wave speed is quadratic in $r$, i.e. $c(r)=\frac{(1-r^2)}{2}c_0$. 

\section{Poincar\'e Disk Lattice Tessellations}
\begin{figure}[!ht]
\centering
\includegraphics{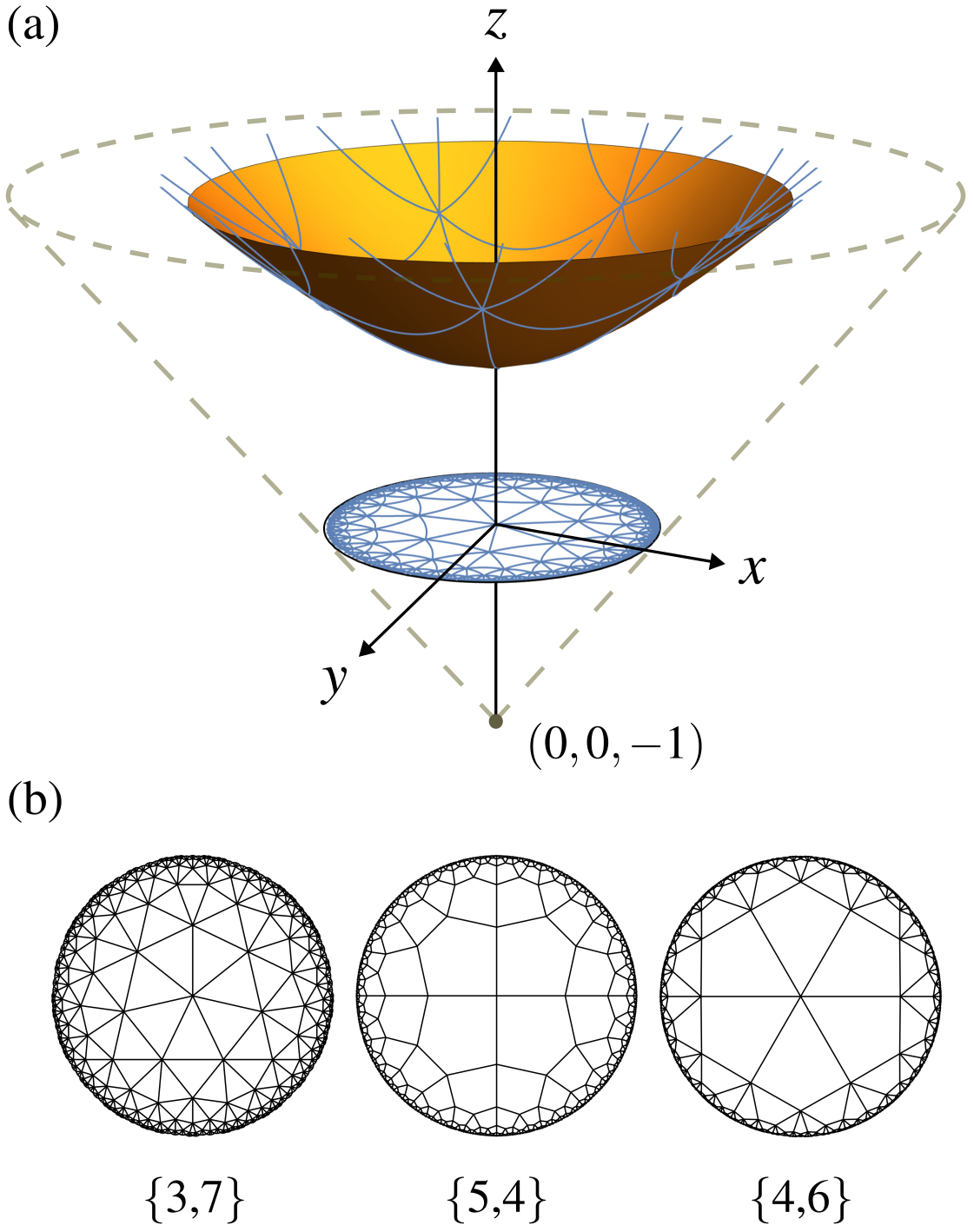}
    \caption{Tessellations of the Poincar\'e disk model of hyperbolic space. (a) A hyperboloid sheet (orange) tessellated by regular triangles with vertex coordination number 7 projected onto the flat unit disk (blue), illustrating a \{3,7\} tessellation of the Poincar\'e disk model of hyperbolic space. (b) Example hyperbolic lattices with sites defined by vertices of tessellations of the Poincar\'e disk connected by straight Euclidean line segments. The tessellations are denoted by their Schl\"afli symbols, where the first number defines the number of sides of each regular polygon and the second number defines the number of polygons shared by each vertex. The \{3,7\} lattice of (a) is shown in (b) from a top down view.}
    \label{Fig2}
\end{figure}

The properties of a system governed by an equation similar to Eq.~\eqref{hyperbolic wave eq} may be investigated by considering a hyperbolic tessellation in the Poincar\'e disk as illustrated in Fig.~\ref{Fig2}a, which shows a regular triangular tessellation on the hyperboloid and its resulting projection on the disk. Figure~\ref{Fig2}b shows other sample tessellations, obtained by connecting hyperbolically distributed lattice sites with straight Euclidean ligaments. Three different lattices are shown, distinguished by their Schl\"afli symbols $\{p,q\}$ where $p$ is the number of sides of the tessellating polygon and $q$ is the vertex coordination number. A hyperbolic polygon, or tile, hosts an interior angle sum less than $(p-2)\pi$ due to its distortion in curved space, giving the governing constraint $(p-2)(q-2) > 4$ ~\cite{beardon1979hyperbolicP,greenberg1993euclidean}. This rule is satisfied by an infinite ensemble of $\{p,q\}$ sets, granting a limitless number of possible tessellations. In contrast, the Euclidean plane requires that tessellations satisfy the equality $(p-2)(q-2) = 4$, a constraint met by only three possibilities: \{3,6\}, \{4,4\}, and \{6,3\}. We form the hyperbolic tessellations by first nucleating a seed tile at the origin and then repeatedly reflecting that tile about its sides until the desired tile count is reached. Each group of tile reflections surrounding the previous group marks a new generation, where an infinite number of generations is required to tile the full Poincar\'e disk~\cite{ruzzene2021dynamics}. The vertex count of hyperbolic tessellations grows exponentially as a function of generation~\cite{baek2009percolation,gu2012crossing,zhang2022observation,stegmaier2022universality,crystallographyhyp,cheng2022band}. As a result, all hyperbolic tessellations produce highly connected boundaries independent of $p$ and $q$, and we find that different tilings produce similar spectral properties. The following investigations focus on the \{3,7\} tessellation, which is selected as a representative of properties also found in other tessellations.

Figure~\ref{Fig2} illustrates how even though the edges of regular triangles in hyperbolic space are all equal, in the hyperboloid and disk they are distorted. The challenge, then, is how to account for these distortions so that tessellations built in Euclidean space can still exhibit hyperbolic characteristics. One approach is to build a system where these lengths are unimportant. In this scenario, continuous dynamics are recovered because the graph Laplacian of the system (which depends only on the lattice connectivity pattern, described as a graph) reduces to the continuous Laplace-Beltrami operator in the long-wavelength limit~\cite{lenggenhager2022simulating,boettcher2020quantum}. An alternative approach is to increase the effective distances between lattice vertices by stretching the paths connecting these sites. For example, when these paths are made to no longer follow the system's geodesics, they can be carefully arranged so that they are of equal lengths while maintaining the same vertex connectivity pattern~\cite{kollar2019hyperbolic}. This observation is particularly relevant for this study, which considers the mechanical implementation of these lattices, wherein interactions are dependent upon the Euclidean distance between connected sites. This distances the tessellations from the continuous hyperbolic medium description of Eq.~\eqref{hyperbolic wave eq}. However, the results presented below will show how some interesting properties of the system, and specifically the existence of boundary-dominated frequency ranges, are preserved.

\section{Numerical investigation of spectral properties}

We investigate the ability of the \{3,7\} lattice to localize vibrations near its boundary by evaluating its spectral properties and comparing them to those of continuous hyperbolic and Euclidean disks. Both the hyperbolic and Euclidean disks are modelled via finite element discretizations of the variational formulations of Eq.s~\eqref{Euclidean wave eq},\eqref{hyperbolic wave eq}. Using the weak form module in the COMSOL environment, these statements are approximated by Galerkin forms whereby the problem is defined over a discrete space of admissible trail and weighting functions. This allows us to assemble a finite element matrix formulation over a discrete mesh to approximate the eigenvalue problem. We partition each disk into 2880 quadrilateral elements via an unstructured mesh with a maximum element size of $1/25$ the disk radius to ensure proper convergence. We employ Hermite cubic shape functions for the element displacement interpolation. The eigenvalue problem is solved for the first 600 modes.

The \{3,7\} lattice is modelled as a network of 1D Timoshenko beams undergoing strictly out-of-plane vibrations. We discretize each beam into a finite element mesh with a maximum element length of 1/10 the shortest lattice beam length, which is expected to ensure the accurate capture of the first 600 modes. The resulting mesh comprises 3024 1D elements which again employ Hermite cubic shape functions for the element displacement interpolation. The beams are made of 6061 aluminum (Young's Modulus 70 GPa, density 2700 kg/m\textsuperscript{3}, Poisson's ratio 0.33) with cross-sectional width 3.81 mm (0.15 inches) and height 4.83 mm (0.19 inches). Constrained by the small features that arise near the rim of Poincaré disk tessellations, we study a three-generation version of the lattice, holding 85 lattice vertices which fill the disk to radial coordinate $r\leq 0.918$. This is scaled to a physical radius $R=$ 150 mm. These dimensions reflect the experimental geometry described in the upcoming sections (see Figures~\ref{Fig4}a,b and~\ref{Fig5}). The summarized design provides a proper balance between the computational cost required for analysis and the ability to retain the key properties under investigation. All considered systems have free boundary conditions.

In analyzing the spectral properties, the eigenvalues are characterized in terms of an integrated density of states, while the eigenstates are classified on the basis of a localization index which quantifies the degree of spatial localization near the disk boundaries. We express this integrated density of states (IDS) \cite{bellissard2001noncommutative,carmona2012spectral,liu2022topological,pal2019topological} as a function of modal index, $i$, by

\begin{equation}
    IDS(i) = \displaystyle  \frac{\#\{n \, | \, f_{n} \leq f_i\}}{N},
    \label{IDS}
\end{equation}

\noindent where $\#$ is a counting operator returning the cardinality $n$ of the set of eigenfrequencies available below the $i^{th}$ eigenfrequency $f_i$, while $N$ denotes the total number of eigenfrequencies considered. Degenerate eigenfrequencies are repeated according to their multiplicity. The IDS naturally lies in the range $[0, 1]$ as it is normalized by $N$. We here consider the first $N=$ 600 states of each system, which captures the IDS within a frequency range corresponding to wavelengths of dynamic deformation on the order of the smallest lattice beam length. Our choice of truncating the lattice to the third generation implicitly limits the occurrence of such wavelengths and therefore informs a suitable upper bound to $N$. 

We classify each state by estimating the extent to which it is localized towards the domain center or outer boundary. This classification is based on the computation of a \emph{localization index}, adapted from the index first applied to discrete hyperbolic lattices \cite{ruzzene2021dynamics} and defined as 

\begin{equation}
	\mathcal{L}_i = \frac{\iint \, (\frac{r}{R})^2 u_i^{2}(\bm r) \, dA}{\iint \, u_i^{2}(\bm r) \, dA}.
\label{localization factor}
\end{equation}

Here $r$ is the magnitude of the position vector $\bm r=r(\cos{\theta}\hat{i}+\sin{\theta}\hat{j})=r\hat{r}$ with origin at the center of the domain, $u_i(\bm r)$ is the scalar displacement field of the $i^{th}$ eigenstate being classified, and $R$ is the radius of either disk or lattice. The localization index is the mean squared (normalized) radial position of the squared displacement field. Simply put, it is a weighted average of the squared radial position $(\frac{r}{R})^2$, where the weight $u_i(\bf r) ^2$ emphasizes locations along the radius with significant displacements. The square root of $\mathcal{L}_i$ provides an effective nondimensional radius between 0 and 1 where we find the majority displacement, on average. A value close to 1 indicates the state is predominantly localized near the boundary (i.e. $r\rightarrow R$), while a value near 0 indicates localization near the origin ($r\rightarrow0$). 

Equation ~\eqref{localization factor} is a high level description of $\mathcal{L}_i$ acting on continuous domains. As we discretize our domains into finite elements, we compute the localization index numerically as

\begin{equation}
     \mathcal{L}_i = \frac{\sum_{e=1}^{n_{el}}\int_{\Omega^{e}}(\frac{r}{R})^2 \bra{N^e(\bf{r})}\ket{d^e}^2 d\Omega }{\sum_{e=1}^{n_{el}}\int_{\Omega^{e}}\bra{N^e(\bf{r})}\ket{d^e}^2 \, d\Omega},
     \label{loc numerical}
\end{equation}

\noindent where our displacement field is now expressed as the inner product $u_i(\textbf{r}) = \bra{N^e(\bf{r})}\ket{d^e} =\sum_{a=1}^{n_{en}}N^e_a(\bm r)d^e_a$ where $\bra{N^e(\bf{r})}$ is the vector of Hermite cubic shape functions and $\ket{d^e}$ is the vector of nodal displacements, each of length $n_{en}$, the number of element nodes. This is the element interpolation of the displacement at position vector $\bf{r}$ within the element domain $\Omega^e$ based on the computed nodal displacements. The integration is carried out numerically in the material (Lagrangian) frame via 7-point Gaussian quadrature giving exact solutions for our polynomial integrands which are at most order 12. We integrate over each individual element domain $\Omega^e$ so as to integrate over the entire discretized domain, given by $\Omega^h = \textbf{A}_{e=1}^{n_{el}}\Omega^e $ with assembly operator $\bf{A}$~\cite{hughes2012finite} acting over the number of elements $n_{el}$. We then sum these integrals in both the numerator and denominator over $n_{el}$ (2880 for the disks and 3024 for the lattice), taking the ratio as the localization index. 

Through $\mathcal{L}_i$, we classify each mode as either interior-dominated or boundary-dominated by setting a decision boundary, or threshold, $\mathcal{L}_t=0.5$. This value is chosen because it corresponds to a displacement field averaged to a normalized radial position of $(\frac{r}{R})_t = \sqrt{0.5} \approx 0.707$, which is the radius that exactly separates equal areas between the interior and boundary regions of the disks or lattice. Thus, $\mathcal{L}_i>\mathcal{L}_t$ indicates high modal activity in the boundary region while $\mathcal{L}_i<\mathcal{L}_t$ indicates high modal activity in the interior region. Values near $\mathcal{L}_t$ describe states with global behavior, which are ultimately classified relative to $\mathcal{L}_t$ for simplicity.

\begin{figure}[h!]
    \centering
    \includegraphics{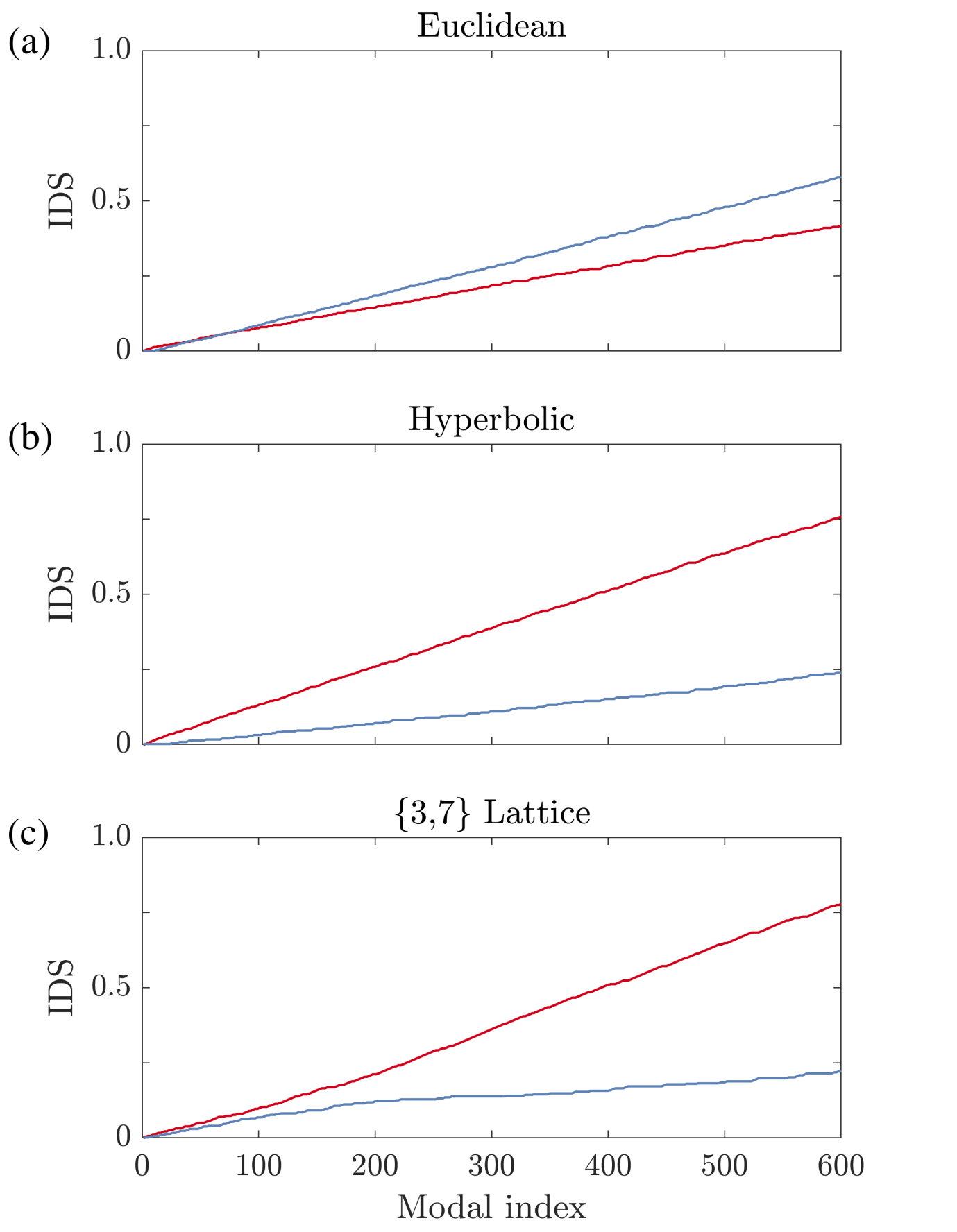}
    \caption{IDS curves versus modal index number for (a) a Euclidean disk, (b) a hyperbolic disk, and (c) the \{3,7\} lattice. The  blue IDS curves correspond to interior-dominated (blue) states, while the red curves are for boundary-dominated ones.  The Euclidean spectrum begins with a similar number of interior (blue) and boundary (red) states and becomes dominated by interior modes for increasing mode indices, while the hyperbolic disk shows a dominance of boundary modes. The \{3,7\} lattice holds an approximately equal proportion of interior and boundary modes until around the 80th mode, where the boundary mode count begins to dominate and ultimately give a proportion similar to the hyperbolic disk.}
    \label{Fig3}
\end{figure}

After applying $\mathcal{L}_t$ to separate the spectra into subsets of boundary and interior states, we evaluate their respective, separate IDSs. In Fig. \ref{Fig3}, we plot the boundary and interior IDS of each system, coloring the curves corresponding to boundary states red and interior states blue. The horizontal axis marks the total states plotted among both red and blue curves up to a modal index such that the sum of both corresponding IDSs values gives the proportion of the 600 modes considered. Evidently, by the 600th mode, the sum of bulk and boundary IDS is 1. The curves for the Euclidean and hyperbolic continuous disks governed by Eq. \eqref{Euclidean wave eq} and Eq. \eqref{hyperbolic wave eq} respectively are shown in Fig.s~\ref{Fig3}a,b. The Euclidean disk (Fig.~\ref{Fig3}a) shows an overall higher number of interior modes than boundary modes, as indicated by the blue curve ultimately exceeding the red one. The opposite happens for the hyperbolic disk which has a consistently and significantly higher boundary IDS than interior IDS (Fig.~\ref{Fig3}b). This is due to two features, highlighted in other studies of Poincar\'e disk wave dynamics~\cite{ruzzene2021dynamics,lenggenhager2022simulating}. The first is that mode shapes on the Poincaré disk are similar to those of the Euclidean disk, but with more prominent localization toward the outer rim. The second is that more boundary modes populate lower frequencies in the hyperbolic Poincar\'e disk than in the Euclidean disk. This can be understood by considering that hyperbolic distances are most distorted near the disk's boundary, where they are longer than their Euclidean counterparts. As such, wavelengths of angular, boundary-localized waves in the Poincar\'e disk are longer than those more centrally located. This is directly deduced from Eq. \eqref{metric} in geodesic polar coordinates with curvature $K=-1$: $d\rho^2 = dr^2 + \sinh^2{(r)} d\theta^2$ ~\cite[pg. 133]{berger2003panoramic}. We consider the path of a boundary-localized wave along an origin-centered circle parametrically described by $r(t)=R$ and $\theta(t) = t$. The hyperbolic circumference $C_H$ of this circle is given by the arc length in one revolution, computed from the metric by $C_H = \int_{0}^{2\pi}\sqrt{(\frac{dr(t)}{dt})^2 + \sinh^2{(r(t))} (\frac{d\theta(t)}{dt})^2 }dt$. We directly find that the hyperbolic circumference is $C_H = 2\pi\sinh{(R)}$, which grows superlinearly in radius by $\frac{dC_H}{dR} = 2\pi\cosh{(R)}$. On the other hand, the circumference of a Euclidean circle grows constantly by $\frac{d}{dR}2\pi R = 2\pi$. Therefore, boundary modes in the Poincar\'e disk with angular oscillations at radial point $R>0$ have longer circumferential wavelengths than Euclidean boundary modes at the same $R$ with the same number of angular zeros. Thus, boundary modes in the Poincar\'e disk have lower frequencies and hence lower modal indices than similar ones in the Euclidean disk.

By conducting a similar analysis on the \{3,7\} lattice, we can ascertain to what extent such a lattice captures the spectral signature of the hyperbolic plane. The lattice's IDS shown in Fig. \ref{Fig3}c reveals a spectral dominance of boundary states in the \{3,7\} lattice similar to that of the hyperbolic disk. While the IDS of interior and boundary modes is approximately balanced at the lower modal indices, a tendency for modes to localize along the boundary is observed to begin around the 80th mode. Below this region, in the start of the spectrum where wavelengths are long relative to lattice beam lengths, the lattice behaves more like the Euclidean domain, with a similar proportion of interior and boundary modes. In contrast, as the characteristic wavelengths decrease, we observe the surfacing of features similar to those of the hyperbolic disk, whereby a larger proportion of boundary modes in the lattice exists. This demonstrates the ability of the considered lattice to mimic the boundary-dominated signature of the hyperbolic disk above a certain threshold frequency. At the highest considered mode, the lattice's spectrum is populated by 78\% boundary modes, which is close to the hyperbolic disk's 76\% boundary modes while the Euclidean disk has only about 42\% boundary modes.

\section{Numerical demonstration of edge-confined wave propagation}
\begin{figure*}[!htb]
    \centering
    \includegraphics[width=\textwidth]{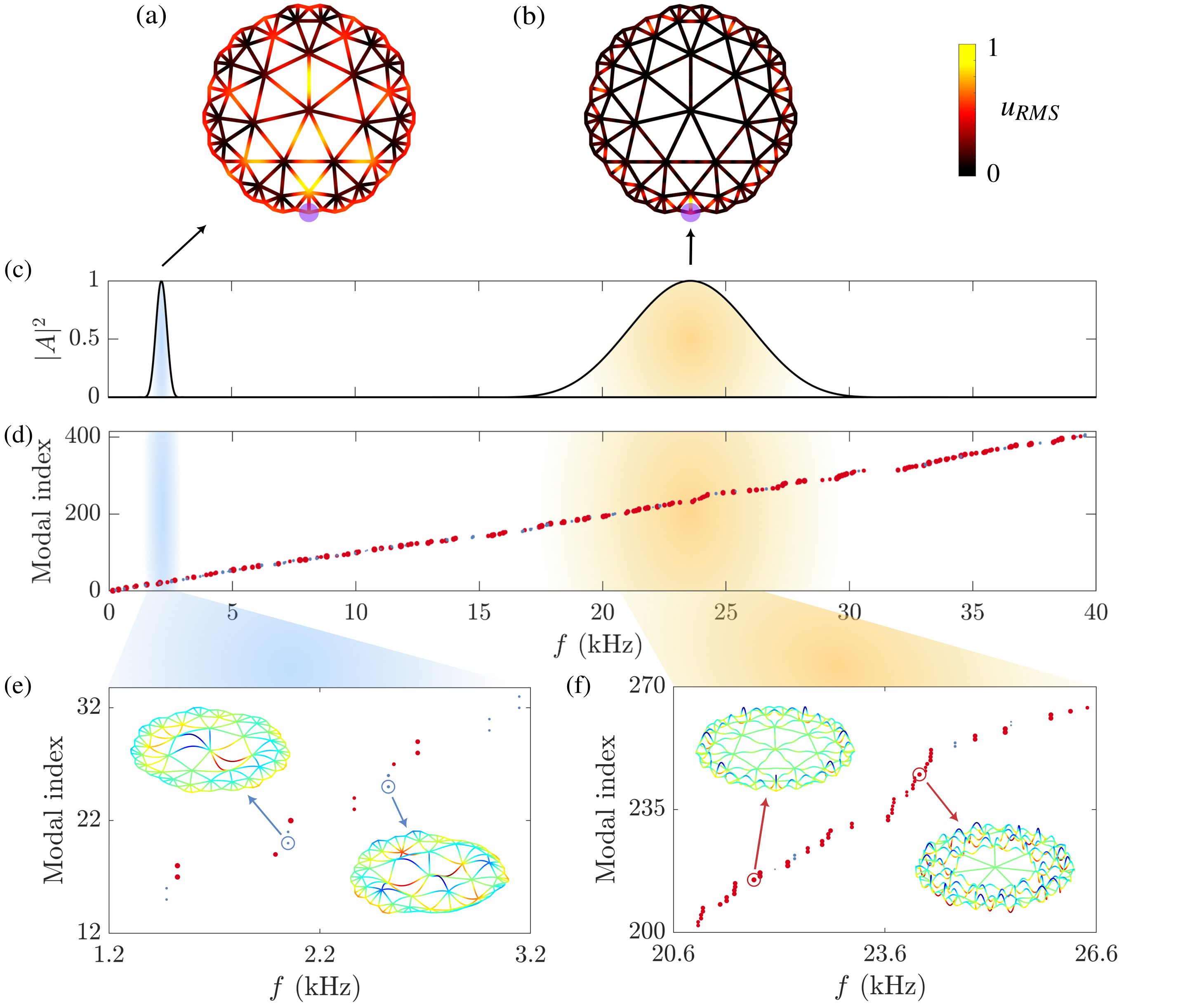}
    \caption{(a,b) Time-averaged squared displacement (RMS) fields of a \{3,7\} lattice excited at its boundary (transparent purple dot) by transient inputs centered at (a) 2.2 kHz and (b) 23.6 kHz. (c) Frequency content of the 2.2 kHz and 23.6 kHz pulses, each with equivalent fractional bandwidths. (d) Spectrum of the \{3,7\} lattice in the 0-40 kHz range: eigenvalues are plotted with red and blue dots based on the associated localization index, which also determines the size of the dot. (e) Zoomed in portion of the spectrum in the frequency range of the 2.2 kHz pulse, highlighting two interior-dominated modes. (f) Zoomed in portion of the spectrum in the frequency range of the 23.6 kHz pulse highlighting two boundary-dominated modes.
    }
    \label{Fig4}
\end{figure*}

Next, we investigate the lattice response to transient vibratory inputs to understand how it is influenced by its boundary-dominated spectrum. We do so via time-domain simulations of the lattice response when subjected to broadband pulse excitations applied to its boundary. These simulations are carried out using the COMSOL finite element solver with the same geometry, parameters, boundary conditions, and mesh previously used for the eigenfrequency study. The excitation is a Hanning-modulated sine wave including six cycles of the signal's center frequency. The duration of each simulation is 12 times the duration of the excitation, which is equal to the number of cycles times the period of the excitation center frequency. Figures~\ref{Fig4}a,b show the location of the input excitation and the color plots of the root mean square (RMS) lattice displacement field averaged over the simulation time for broadband inputs centered at 2.2 kHz (\ref{Fig4}a) and 23.6 kHz (\ref{Fig4}b). The RMS displacement fields are normalized by the maximum RMS value according to the reported color bar, which shows a variation in intensity from black (0) to yellow (1). In Fig.~\ref{Fig4}a, the response amplitude is globally distributed with the highest displacements near the excitation point and in the center. In Fig.~\ref{Fig4}b, the response amplitude is significantly larger at the boundary with virtually no motion in the bulk, confirming the boundary-governed spectral properties of the lattice at high frequencies.

This observation is supported by the spectral content of the inputs, plotted in Fig.~\ref{Fig4}c as normalized power spectral densities. The bandwidths of the two considered frequency inputs are highlighted by shaded blue and gold areas which are projected onto Fig.~\ref{Fig4}d, the lattice's spectrum in the 0-40 kHz range. Each mode in the spectrum is represented as a dot, color coded again in red (boundary mode) and blue (interior mode). The size of the dot is also scaled by the localization index $\mathcal{L}_i$, with very large and very small dots indicating high degrees of localization. The highlighted bands aid in visualizing which modes are excited by the applied input and therefore contribute to the lattice's transient response. Fig. ~\ref{Fig4}e,f provide magnified portions of the spectrum within the frequency contents of each pulse. We see that for the 2.2 kHz pulse, there are a roughly equal number of boundary and interior modes, whereas the 23.6 kHz pulse excites significantly more boundary modes. In fact, the boundary:interior mode ratio in Fig. ~\ref{Fig4}e is 10:9 compared to 56:7 in Fig. ~\ref{Fig4}f.  Example mode shapes are inset in Fig. \ref{Fig4}e and Fig. \ref{Fig4}f which show two interior modes and two boundary modes respectively.

\section{Experimental corroboration of spectrum and edge-confined waves}

\begin{figure}
    \centering
    \includegraphics{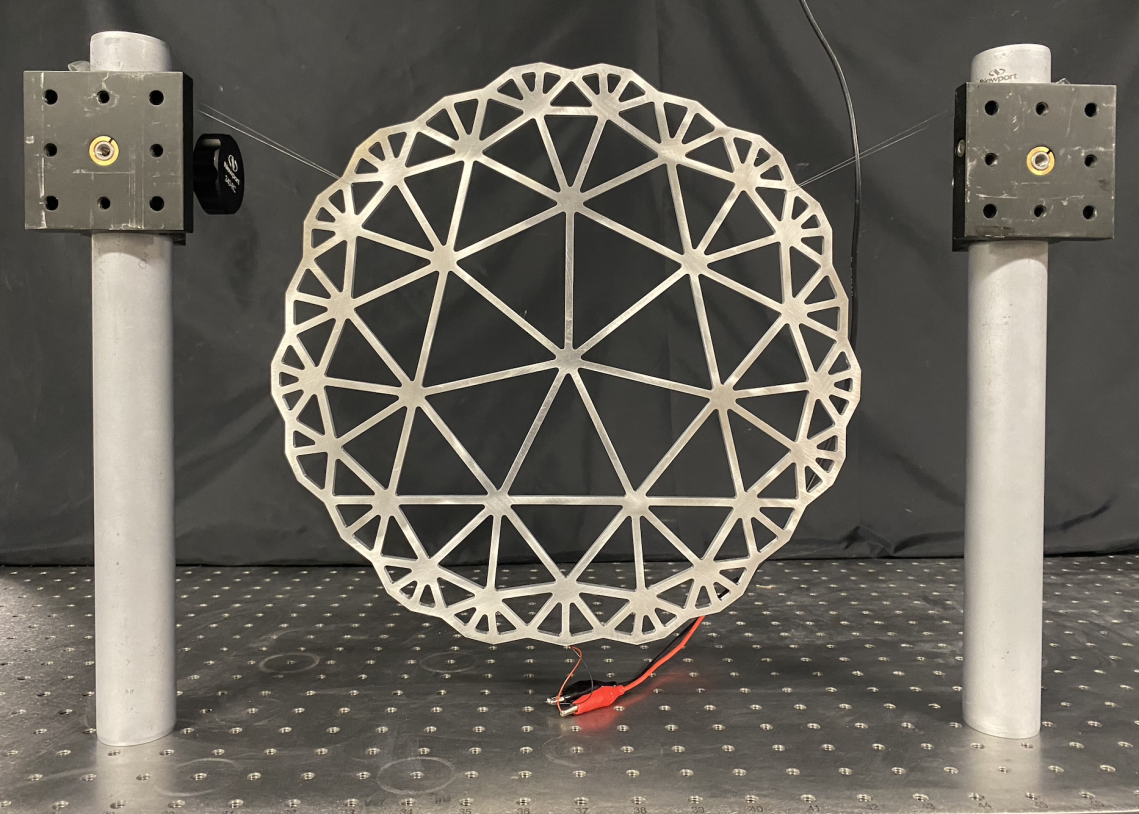}
    \caption{Measurement setup of the experimental \{3,7\} lattice. The lattice, machined out of an aluminum plate, is suspended via nylon strings bound to damped optical posts affixed to an optical table in order to achieve free boundary conditions. A function generator is used to excite the lattice via a piezoelectric transducer at its bottom-most point, where we see red and black alligator clips attached to the leads of this transducer. A scanning laser Doppler vibrometer measures the response of the lattice, and is connected to a data acquisition unit (DAQ) which processes and stores the results.}
        \label{Fig5}
\end{figure}

Transient vibration measurements are conducted on a \{3,7\} elastic lattice to corroborate the spectral properties predicted by the theoretical investigation presented above. The experimental specimen is machined from a single 6061 aluminum plate by milling away triangles. The resulting lattice, pictured in Fig.~\ref{Fig5}, comprises beams of width 3.81 mm and thickness 4.83 mm, as in the numerical model. The overall lattice radius is $R =$ 150 mm. In these tests, the lattice is suspended vertically by thin nylon threads bound to optical posts mounted on an optical table. The lattice faces a laser Doppler vibrometer that takes the vibration measurements, and it is excited by a piezoelectric disc bonded at the boundary site corresponding to the excitation point numerically considered and depicted in Fig.~\ref{Fig4}a. 

First, the modal content of the lattice is evaluated via a frequency domain experiment. Here, the affixed piezoelectric transducer excites the lattice with pseudorandom noise containing frequencies from 0-40 kHz, which is the range numerically studied in the previous section. These measurements highlight the resonance frequencies of the lattice and allow for the estimation of dynamic deflection shapes related to the eigenstates of the system. They are used to compute correlation with numerical mode shapes and to corroborate the corresponding localization indices. Second, transient vibration measurements are conducted to verify the edge-dominated transient response of the lattice when excited by broadband signals in the boundary-dominated high frequency regime compared to the bulk, interior-dominated low frequencies.

\begin{figure}[!htb]
    \centering
    \includegraphics{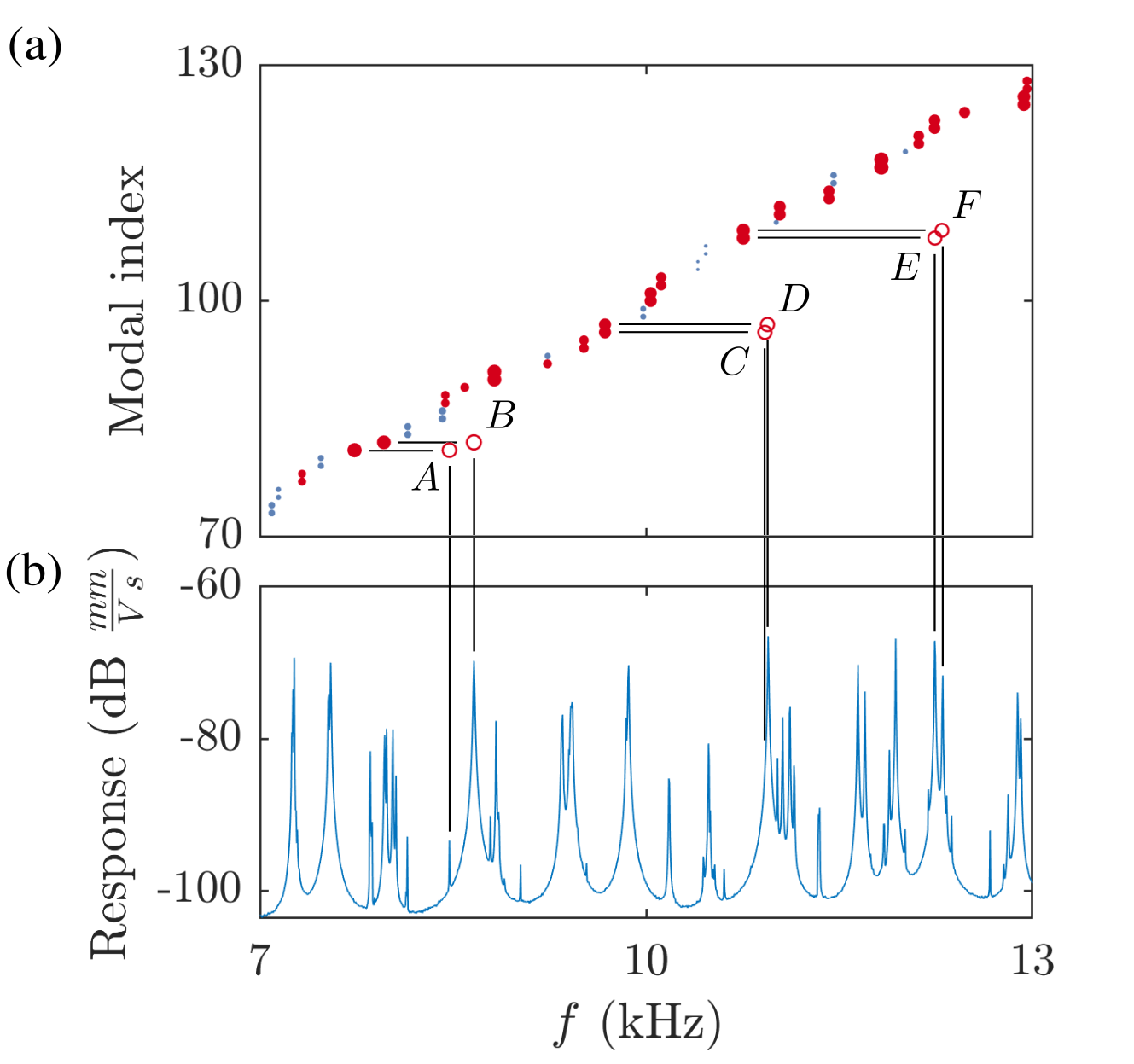}
    \caption{(a) Numerical spectrum of the \{3,7\} lattice in the 7 to 13 kHz range. Modes are ordered by frequency and sized according to their localization index. Solid red dots indicate boundary modes while solid blue dots indicate interior modes. Selected experimental boundary modes are plotted in hollow red circles. Labels A-F indicate the six pairs of modes being compared. (b) Experimental frequency response function in the 7 to 13 kHz range.}
        \label{Fig6}
\end{figure}

In the frequency domain study, the lattice's response is obtained by averaging the responses at each scan point, producing the averaged frequency response in Fig. \ref{Fig6}b. The plot is limited to the 7-13 kHz range, which contains the portion of the spectrum where the IDS begins to become boundary-dominated; thus, it contains the lower frequency boundary modes which can be captured in experiments with less noise. Figure~\ref{Fig6}a maps the measured resonant peaks to those measured experimentally, providing a visual comparison with the numerically calculated spectrum. Each of these numerical-experimental mode pairs is labeled with the letters A-F for tracking purposes. Table~\ref{Table1} compares the numerical eigenfrequencies and localization indices to the experimental resonance peak frequencies and localization indices for pairs A-F, showing an overall good agreement. Some of the discrepancies can be attributed to the fact that the finite size of the CNC router bit produces some rounding of the triangle vertices in the milled sample, which are not included in the numerical description of the lattice. This is expected to lead to local stiffening effects at the vertices relative to the idealized numerical model, which may systematically augment experimental frequencies compared to numerical ones.

\begin{table}
    \centering
    \renewcommand{\arraystretch}{1.3}
    \setlength{\tabcolsep}{5pt} 
    \begin{tabular}{|c|c | c|c | c|} 
        \hline 
        Mode & \multicolumn{2}{c|}{Frequency [kHz]} & \multicolumn{2}{c|}{Localization Index} \\ \cline{2-5}
        & \multicolumn{1}{c|}{Numerical} & \multicolumn{1}{c|}{Exp.} & \multicolumn{1}{c|}{Numerical} & \multicolumn{1}{c|}{Exp.} \\ \hline
        A & 7.73 & 8.47 & 0.87 & 0.83\\ 
        B & 7.96 & 8.66 & 0.85 & 0.87\\ 
        C & 9.68 & 10.92 & 0.75 & 0.81\\ 
        D & 9.68 & 10.94 & 0.75 & 0.81\\ 
        E & 10.75 & 12.24 & 0.82 & 0.80\\ 
        F & 10.75 & 12.29 & 0.82 & 0.79\\ \hline
    \end{tabular}
    \caption{Natural frequencies and localization indices of six modes (A-F) in numerical simulations and experimental (``Exp.") measurements.}
    \label{Table1}
\end{table}

\begin{figure}[!htb]
    \centering
    \includegraphics{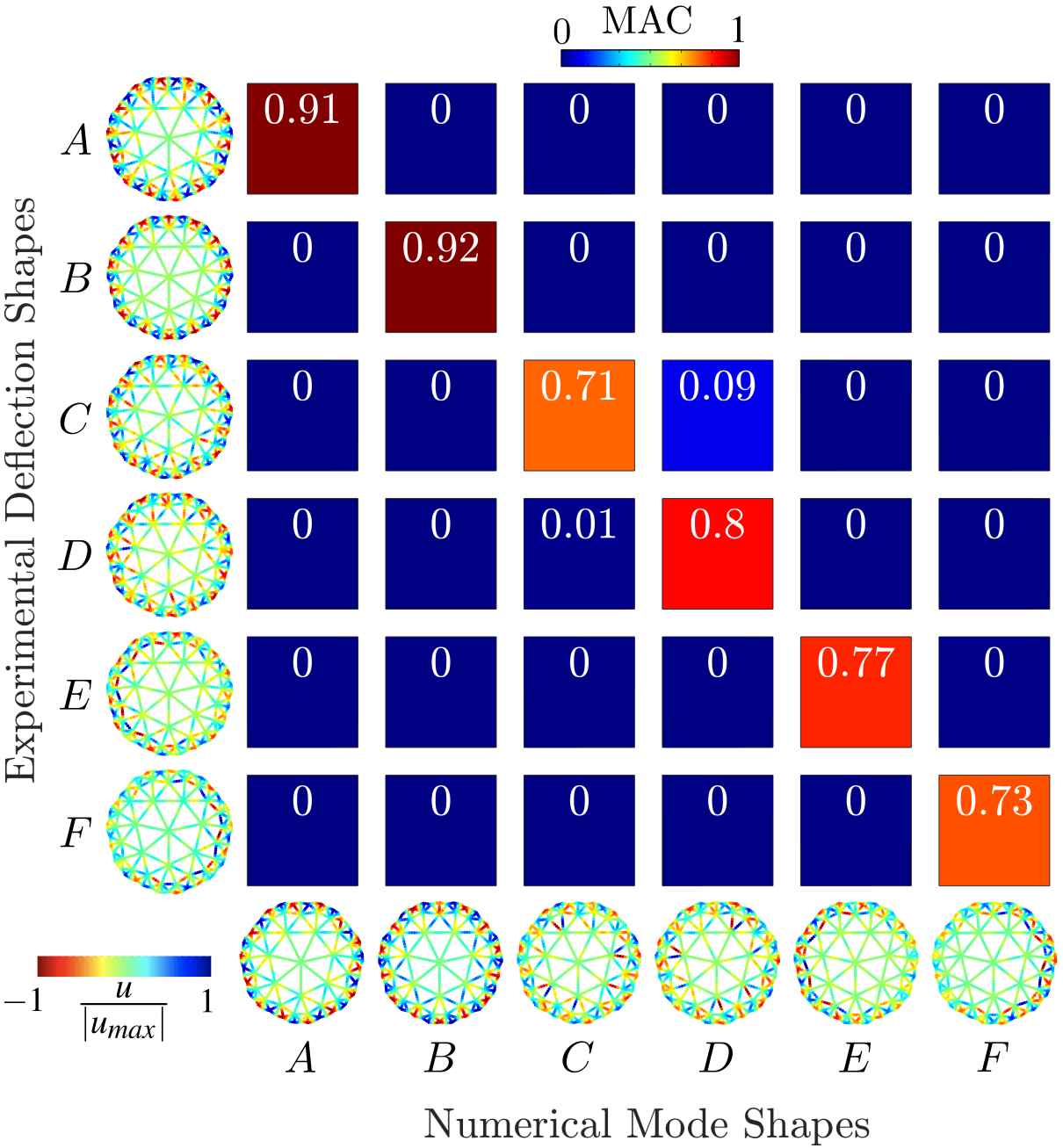}
    \caption{Modal Assurance Criterion matrix comparing measured deflection shapes to numerically calculated mode shapes. (left) Six experimentally obtained operational deflection shapes of manufactured 6061 aluminum \{3,7\} elastic hyperbolic lattice. (bottom) Six numerical mode shapes of the modelled \{3,7\} elastic hyperbolic lattice.}
        \label{Fig7}
\end{figure}

The experimental response provides the necessary information to construct the operational deflection shapes of the lattice based on measured velocities at scanned points for a given frequency peak. These operational deflection shapes are plotted along with numerically obtained mode shapes in Fig.~\ref{Fig7}. By organizing the experimental and numerical shapes into individual mode shape matrices, $\{\phi_e\}$ and $\{\phi_n\}$ respectively, we can further quantify their consistency via the modal assurance criterion (MAC), which provides a matrix with entries $(r,q)$ given by~\cite{pastor2012modal}

\begin{equation}
    MAC(r,q) = \frac{|\{\phi_e\}_r \{\phi_n\}_q|^2}{(\{\phi_e\}_r^T\{\phi_e\}_r) (\{\phi_n\}_q^T\{\phi_n\}_q)},
\end{equation}

\noindent where $\{\phi_e\}_r$ and $\{\phi_n\}_q$ are the individual experimental and numerical mode shapes being compared for the given $(r,q)$ entry. The MAC is effectively a normalized inner product, evaluating to 1 if the measured deflection shape is fully consistent with the numerically calculated mode shape and 0 if they are linearly independent, though a true orthogonality check warrants the inclusion of the mass or stiffness matrix. This evaluation is particularly useful in comparing intricate mode shapes which are challenging to examine with the naked eye, while at the same time providing quantifiable validation of the numerical model. In order to compute the MAC, the numerical and experimental shapes must lie in the same vector space. As such, the numerical mesh is interpolated linearly at the experimental scan points, giving $\{\phi_e\}$ and $\{\phi_n\}$ equal ranks. 

\begin{figure*}
    \centering
    \includegraphics{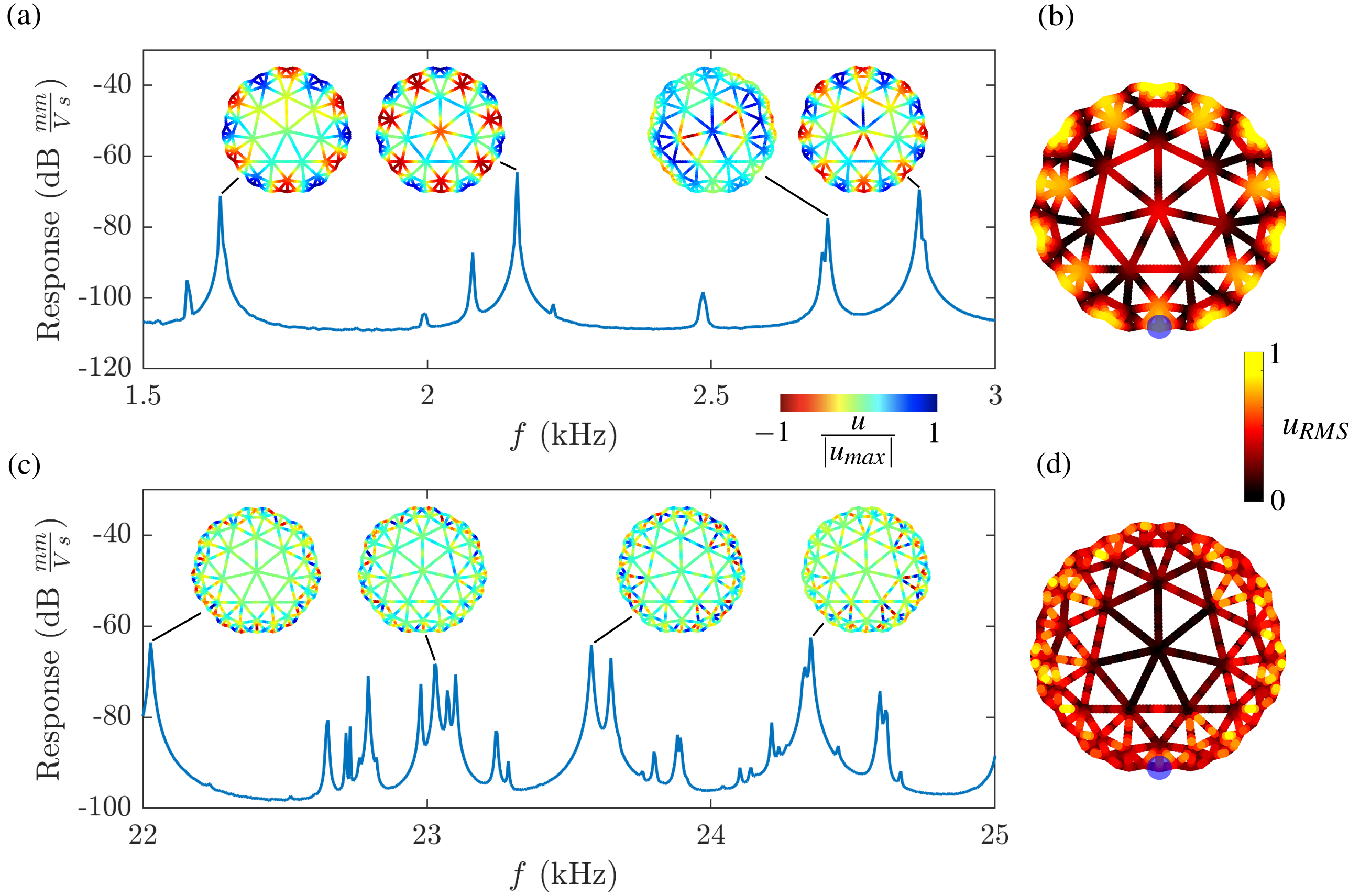}
    \caption{Experimental RMS time-averaged displacement field of the \{3,7\} lattice in response to (b) a 2.2 kHz pulse in the low frequency regime, and (d) a 23.6 kHz pulse in the high frequency regime, both excited through a piezoelectric transducer affixed to an edge node (marked with purple dots). Averages are taken over the course of 12 times the duration of each excitation. Corresponding experimental frequency responses within the frequency content of each pulse are shown in (a) and (c). The four experimental deflection shapes with the highest response peak are inset in each response spectrum.}
    \label{Fig8}
\end{figure*}

To confirm the time-dependent simulations, we suspend and excite the lattice by the same nylon threads and piezoelectric transducer as before. The transducer, affixed at the marked purple dot in Fig.~\ref{Fig8}a, excites the lattice with the 2.2 kHz and 23.6 kHz pulses from numerical simulations. A scanning laser Doppler vibrometer measures out-of-plane displacements across the entire lattice. The RMS displacement fields averaged over the course of 12 times the duration of each excitation are plotted in Fig.~\ref{Fig8}b and ~\ref{Fig8}d where they are normalized to a maximum of 1. Accompanying these averages in Figs. ~\ref{Fig8}a,c are the experimental frequency responses in the neighborhood of each pulse's center frequency. Inset in each figure are the four experimental deflection shapes with the highest response amplitude. We see that for the 2.2 kHz pulse, the four highest peaks correspond to one boundary-dominated mode and three global modes. For the 23.6 kHz pulse, we see that the four highest peaks correspond to four boundary modes. As such, the anticipated global RMS from Fig. ~\ref{Fig8}b and boundary-dominated RMS in Fig. ~\ref{Fig8}d are observed. Compared to the numerical study, we see more boundary activity in experiments overall for both the low and high frequency RMS displacements. We anticipate this is due to the systematic upward shift of experimental frequencies, and thus the inclusion of a additional higher frequency modes in the excitation frequency contents. Nonetheless, the high frequency pulse excites virtually zero activity in the lattice center which encapsulates the entire first generation of beams. This feature could find application in protecting bulk media from vibrations incident on the boundary. Such edge-states are reminiscent of those in topologically protected systems.

\section{Conclusions}

In this paper we report numerical and experimental observations of the boundary-dominated nature of vibrations in a 2D elastic hyperbolic lattice. We first numerically compute the spectrum of the hyperbolic lattice compared to Euclidean and hyperbolic continuous disks, and show that the straight-edged hyperbolic lattice retains some of the boundary dominated features of hyperbolic media, in particular at high mode numbers. This spectrum is experimentally confirmed via frequency domain experiments using a laser Doppler vibrometer. We then numerically investigate the response of the lattice to boundary-incident pulses which reveals its capability to confine incident vibrations to its edge for excitations with high enough frequency content. Examples are provided for pulses centered at 2.2 kHz and 23.6 kHz, which highlight the boundary-dominated behavior in the high frequency where vibrations tend to excite the smaller intricacies of the lattice located at the boundary.
Experimental measurements are then conducted to confirm the numerical observations and specifically to verify the boundary dominated nature of the vibratory response in the high frequency. By leveraging the boundary-dominated spectrum of the hyperbolic plane, the considered hyperbolic lattice acts as a host for edge-confined states reminiscent of topologically protected systems, some of which have recently been uncovered in hyperbolic space. This work presents a feasible way to study hyperbolic lattices where couplings between sites are not all equal, allowing for a natural and simple experimental platform to study elastic waves in hyperbolic lattices. As such, it may allow for the realization of as yet largely unexplored hyperbolic metastructures that are deployed in various geometrical domains as well as applications in architected materials with broadband vibration confinement capabilities.

\section{Acknowledgements}

This work is supported by NSF award number 2131758. The authors declare no competing interests.

\bibliographystyle{unsrt}
\typeout{}
\bibliography{bibliography}

\end{document}